\documentclass[preprint]{aastex}

\begin{document}

\title{Five new {\it{INTEGRAL}} unidentified hard X-Ray sources uncovered by \textit{Chandra}}

\author{M. Fiocchi\altaffilmark{1},
L. Bassani\altaffilmark{2},
A. Bazzano\altaffilmark{1},
P. Ubertini\altaffilmark{1},
R. Landi\altaffilmark{2},
F. Capitanio\altaffilmark{1},
A. J. Bird\altaffilmark{3}
\altaffilmark{1}{INAF/IASF-Roma, Via Fosso del Cavaliere 100, I-00133, Roma, Italy}\\
\altaffilmark{2}{lASF/IASF-Bologna, Via P. Gobetti 101, I-40129 Bologna, Italy}\\
\altaffilmark{3}{School of Physics and Astronomy, University of Southampton, SO17 1BJ, Southampton, UK}
}

\begin{abstract}
The IBIS imager on board {\it{INTEGRAL}}, with a sensitivity better than a mCrab in deep observations and a point source location
accuracy of the order of few arcminutes, has localized so far 723 hard X-ray sources in
the 17--100 keV energy band, of which a fraction of about 1/3 are still unclassified.
The aim of this research is to provide sub-arcsecond localizations of the unidentified sources, necessary to pinpoint the optical 
and/or infrared counterpart of those objects whose nature is so far unknown. 
The cross-correlation between the new IBIS sources published within the fourth {\it{INTEGRAL}}/IBIS Survey catalogue and the
{\it{CHANDRA}}/ACIS data archive resulted in a sample of 5 not yet identified objects.
 We present here
the results of {\it{CHANDRA}} X-ray Observatory observations of these five hard X-ray sources  discovered
by the {\it{INTEGRAL}} satellite.
We associated IGR J10447-6027 with IR source 2MASSJ10445192-
6025115, IGR J16377-6423 with the cluster CIZA J1638.2-6420, IGR J14193-6048 with
the pulsar with nebula PSR J1420-6048 and IGR J12562+2554 with the Quasar SDSS
J125610.42+260103.5. We suggest that the counterpart of IGR J12288+0052 may be
an AGN/QSO type~2 at a confidence level of 90\%.

\end{abstract}


\section{Introduction}

The International Gamma-Ray Astrophysics Satellite (INTEGRAL, Winkler et al. 2003) has been operating for
seven years and has discovered a large number of new hard X-ray sources at energies $>$17 keV.
With its mCrab sensitivity and point source location accuracy of a few arcminutes, the
IBIS instrument (Ubertini et al. 2003) has localized more than 723 hard X-ray sources in the 17--100 keV 
energy band (Bird et al. 2009);
about 30$\%$ of these objects are still unidentified, unclassified and/or poorly studied in 
the X-ray band. A similar picture is found in the all-sky BAT survey performed with the Swift satellite (Cusumano et al. 2009).
The IBIS localization is often insufficient to unambiguously identify the optical or infrared counterpart of a 
detected source thus hampering its classification. To overcome this problem, the arcsec positioning now 
available with the current generation of X-ray telescopes is required. While an X-ray position with an accuracy of several
arcseconds (as obtained by Swift or XMM-Newton) can lead to the correct optical or IR identification, in many cases
(and especially in the crowded regions of the Galactic plane and center),
the sub-arcsecond position available with {\it{CHANDRA}} is crucial to obtain a
firm identification. Furthermore, the ACIS/\textit{Chandra} instrument allows measurement of the soft X-ray energy
emission of the source, providing information on the spectral shape and
level of absorption that are important in determining the nature of the unidentified object. \\
Classification and knowledge of the nature of these unidentified objects is very important to
the study of several astrophysical questions. In particular, we need to understand if they belong to new classes of objects,
and to define their timing and spectral characteristics.
The construction of a large complete sample of hard X-ray objects belonging to specific object classes
(such as AGNs detected with {\it{INTEGRAL}}, Malizia et al. 2009) will also allow the definition of
 the spatial distribution and the luminosity function of various classes of objects.
With this aim, we have cross-correlated the list of IBIS sources included in the fourth catalog with the ACIS data
archive, and derived a sample of 5 unidentified objects that can be investigated at X-ray energies.

 \begin{table*}
\caption{{\em {\it{CHANDRA}}} Observation Log\label{tab:obs}}
\begin{tabular}{lcccccc} \hline \hline
IGR Name 	&l&b	& ObsID &  Start Time (UTC) &Detector &Exposure Time (s)\\ \hline \hline
IGR J10447-6027 &287.89&-1.31   	&	9495  &		2008-04-24 00:57:01&  ACIS-I    &31360\\
                && 	&		9849  &		2008-04-26 16:31:04  &ACIS-I    &28840\\
 IGRJ12288+0052	&290.30 &63.19	&    	7754  &		2007-03-12 13:22:54  &ACIS-S    &9540\\
IGR J12562+2554	&344.27&88.38	&  	3212	&	         2002-12-04 15:08:35 &ACIS-S            & 28100\\
IGR J14193-6048 &313.44&0.27 	&  	7640  &		2007-06-14 21:29:10  &ACIS-I    &71130     \\
                &&    		&	2792  &		2002-09-16 19:28:28  &ACIS-S    &10060\\
                &&    		&	2794  &		2002-09-22 01:48:17  &ACIS-S   & 10060\\
 IGR J16377-6423&324.59&11.52    	&	1227  &		1999-08-25 09:35:10  &ACIS-I    &12240 \\
                &&    		&	1281  &		1999-08-25 05:45:39  &ACIS-I    &11530\\
\hline

\end{tabular}
\end{table*}

\section[]{Data reduction and image and spectral analysis}
For each source in the sample we extract the IBIS light curve to study any possible variability of the sample. Figure 1 shows these light curves  in the 20--100 energy band and demonstrates the absence of any meaningful count rate variation.

X-ray data obtained with the ACIS instrument in
the 0.5--8 keV energy band have been analyzed for each source in the sample . The log of all observations is given in Table 1,
which reports the galactic coordinates, the observation code (ID), the observation date, the detector configuration  and
the exposure time. We downloaded the most recent versions of the data products processed
at the \textit{Chandra} X-Ray Center using the pipeline ASCDS with versions
spanning from 7.6.9 to 7.6.11. In only one case we have reprocessed the data because the 
pipeline version was lower than 7.6.7. Further processing was done
with the CIAO (\textit{Chandra} Interactive Analysis of Observations) software, version 4.1.2, i.e. the 
same version of CALDB (Calibration Data Base), provided by the \textit{Chandra}
X-ray Center and following the science threads listed on the CIAO
website\footnote{Available at http://cxc.harvard.edu/ciao/.}

\begin{figure*}[t]
\includegraphics[angle=-90, width=1\linewidth]{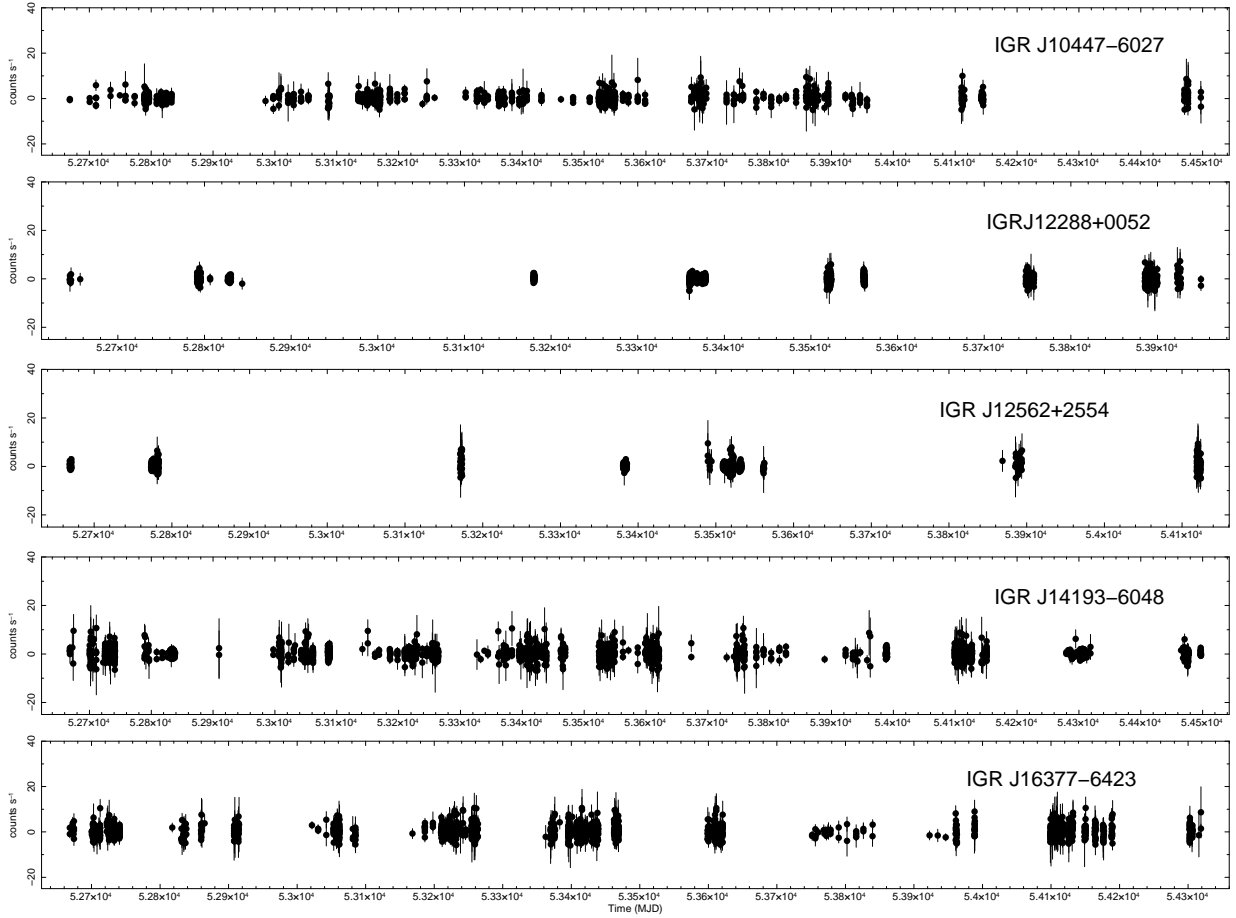}
\figcaption{The {\it{INTEGRAL}}/IBIS litgh curves in the 20-100 keV energy band for each source of the sample
}
\label{fi1}
\end{figure*}

The CIAO routine {\ttfamily wavdetect} was used to search for X-ray sources on the ACIS chips. 
We searched for sources in the 0.5-2 keV, 2-8 keV and 4-8 keV images, setting the threshold
significance for output source pixel list to $9.5\times 10^{-7}$ corresponding to a level
that would be expected to yield only one spurious source.
IGR J16377-6423 is an extended source and hence we used the more appropriate CIAO routine
{\ttfamily vtpdetect} to detect sources within the IBIS error box.

In Table 2,
for each IBIS detection, we report  all the sources detected by ACIS within the IBIS error
circles at 90\% confidence level, their position with relevant error and count rate in three energy bands: 0.5--2 keV, 2--8 keV and 4--8 keV.
We do not take into account the new IBIS error box values reported by Scaringi et al. (2010) since
the improvements in the source positions are only for higher sigma sources (10 sigma or more).
ACIS detections with significance lower than 4 are not reported. Furthermore, we do not discuss here
sources detected only below 2 keV, since they are unlikely to be  associated to the IBIS detections.
The positional uncertainties reported in Table 2,  are 1-$\sigma$ statistical errors calculated by the 
{\ttfamily wavdetect} software. With the spatial resolution limited by the physical size of the
CCD pixels, a systematic pointing uncertainty should be added to this. The pointing 
uncertainties are 0.64 arcsec at 90\% confidence and 1 arcsec at 99\% confidence levels respectively
(Weisskopf, 2005 and available at http://cxc.harvard.edu/cal/).

We next tried to pinpoint the most likely counterpart of the high energy source,
primarily using the source X-ray properties: harder and brighter objects are in fact most likely to
be the true counterpart  of the  IBIS detections and will be discussed in details here. Finally,
using the restricted \textit{Chandra} positions at 99\% confidence level, we
searched for optical/IR counterparts using various on-line archives such as as NED (NASA/IPAC Extragalactic
Database), HEASARC (High Energy Astrophysics Science Archive Research Center) and Simbad (Set of Identifications,
Measurements, and Bibliography for Astronomical Data); the results of
this search are also given in Table 2 but only for sources detected in the 2-8 keV band. 
In this table, we report in bold the coordinates of the best candidate as counterpart for each {\it{INTEGRAL}} source.

{\em {\it{CHANDRA}}} energy spectra have been produced for those counterparts detected with
a significant number of counts in the 4--8 keV energy band (unless otherwise stated).
We extracted source photons from a circular region centered on each source
with an extraction region chosen so that more than 95\% of the energy
was included, using the off-axis angle reported in Table 3.
For the background, we used circular source-free regions in the same CCD as the studied source.
Once the source and background regions were determined,  the CIAO routine {\ttfamily dmextract} was used to
produce energy spectra and {\ttfamily mkacisrmf} and {\ttfamily mkarf} for the response and ancillary files respectively.

For objects with more than one pointing, having checked for the absence of significant
luminosity changes, we performed simultaneous spectral fitting using all the available
spectra. A simple power law model,
absorbed by Galactic absorption, has been used to fit  ACIS spectra, with XSPEC software v11.3.2.
When required by the fit, a power law model with two absorbers has been used:
the first $N_H$  fixed to the Galactic column density (reported in Table 3)
and the second as a  free parameter in order to study the possible presence of intrinsic absorption.
{{The result of this analysis is reported in Table 3 that
lists the column density in excess of the galactic value, the photon index, the 0.3--10 keV flux and the extrapolated 20-40 keV fluxes}}.
The uncertainties on the various parameters are at $90\%$ confidence level for one
parameter of interest ($\Delta\chi^2=2.71$).

\section[]{Notes on individual sources}

In the following section, the result on each individual source is presented.

\subsection{IGR J10447-6027}

\begin{figure}[t]
\includegraphics[angle=-0, width=0.6\linewidth]{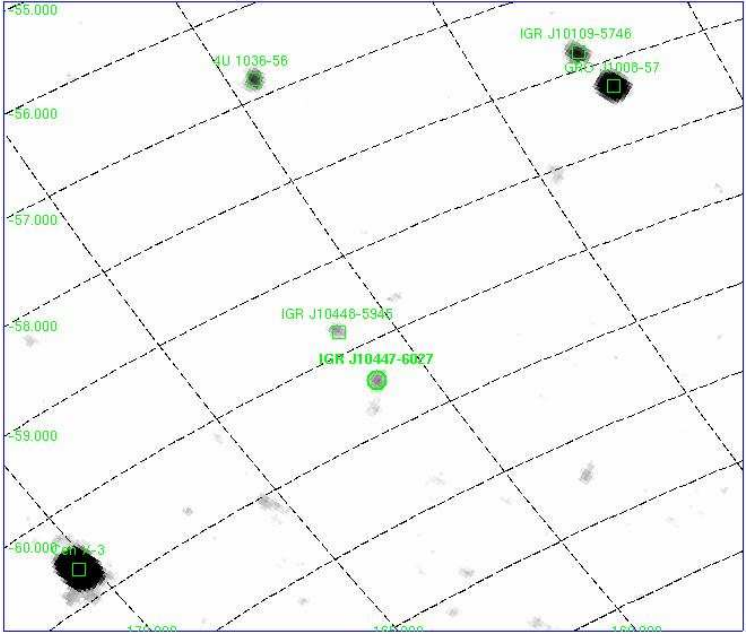}
\figcaption{The {\it{INTEGRAL}}  IBIS/ISGRI mosaic significance image around IGR J10447-6027 in the 18-60 keV energy band for a total exposure of 2189 ks.
}
\label{fi1}
\end{figure}

\begin{figure}[h]
\includegraphics[angle=-90, width=0.8\linewidth]{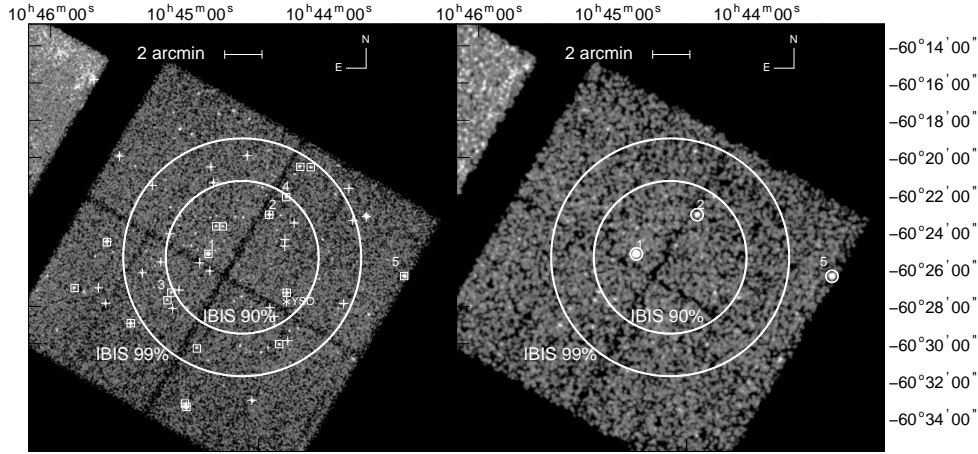}
\figcaption{Left: ACIS 0.5$-$8 keV image of the region surrounding IGR J10447-6027.  
A Gaussian smoothing was applied to the counts distribution with a width of 2 pixels.
The large circles represent the IBIS position and uncertainties expressed as
90\% and 99\% confidence circles (as reported by Bird et al. 2010).
Crosses indicate detections in the 0.5$-$2.0 keV band, squares detections in the 2.0$-$8.0 keV band and
numbers indicate the position of the X-ray sources detected within the
IBIS error box in the hard X-ray band (4$-$8 keV). The asterisk is the position of the YSO, IRAS 104236011.
Right: ACIS 4$-$8 keV image of the region surrounding IGR J10447-6027.
A Gaussian smoothing was applied to the counts distribution with a width of 4 pixels.
The large circles represent the IBIS position and uncertainties
expressed as 90\% and 99\% confidence circles (as reported by Bird et al. 2010).
Circles represent sources detected in the 4$-$8 keV energy band.
}
\label{fig1}
\end{figure}

\begin{figure*}[h]
\includegraphics[angle=-90, width=0.8\linewidth]{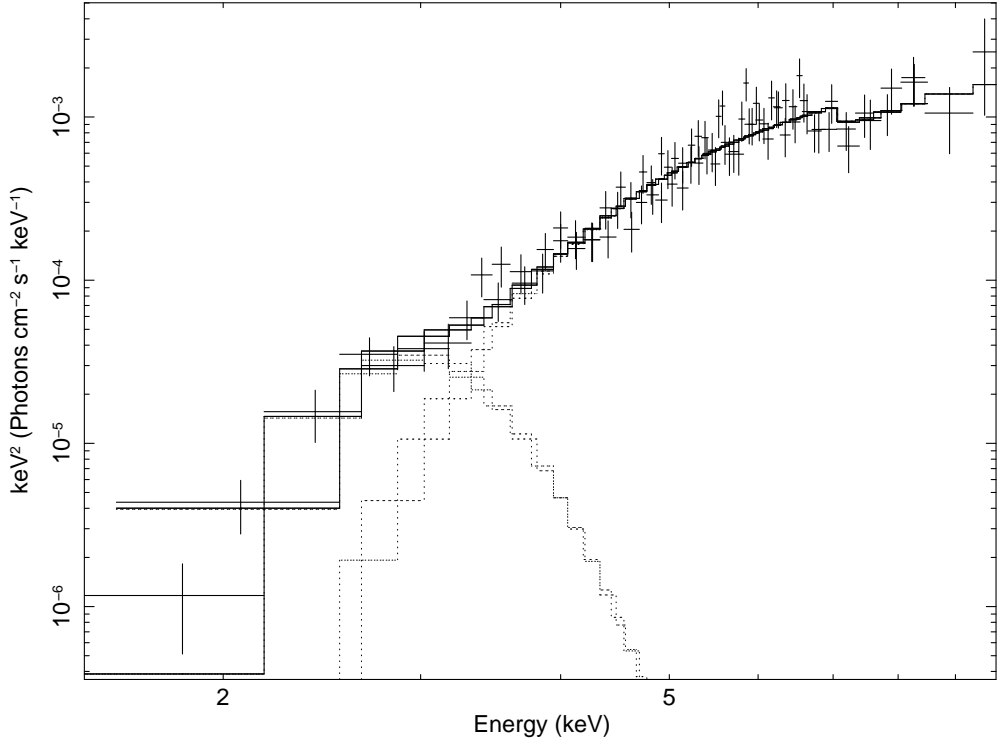}
\figcaption{ACIS spectrum of Source \#1 in the region surrounding IGR J10447-6027, fitted with a simple power law model
plus a thermal BREMS model. }
\label{fig2a}
\end{figure*}

This source was discovered by Leyder, Walter \& Rauw (2008) during the analysis of the region surrounding Eta
Carinae and associated to a young stellar object (YSO,
IRAS 104236011 RA:10 44 17.9; DEC:-60 27 46); it has been included in the IBIS 4$^{th}$ catalog
as a faint persistent object detected in the 18-60 keV energy band at 5.8 sigma level (Bird et al. 2010).
The  IBIS/ISGRI mosaic significance image around IGR J10447-6027 in the 18-60 keV energy band (total exposure of 2189 ks) is shown in Figure 1.
Within the most recent IBIS error box reported by Bird et al.
(2010), we find many soft X-ray sources, 11 objects in the 2-8 keV energy band but only two
in the 4-8 keV energy range (see Figure 2); both objects detected at hard energies fall inside the 90$\%$ IBIS error box.
A bright and hard source (\#5)  is also detected with a signal to noise ratio $\sim7$ at high energies,
but it is outside of the 99\% confidence error circle.
We then performed spectral analysis for the two brightest detections (Source \#1 and \#2).
Both these sources do not show any time variability, in fact in both ACIS pointing the count rates are similar:
Source \#1 has $(2.02\pm0.08)\times10^{-2}$ counts/s and $(2.17\pm0.09)\times10^{-2}$ counts/s
while Source \#2 has $(3.8\pm0.4)\times10^{-3}$ counts/s and $(3.1\pm0.4)\times10^{-3}$ counts/s in the 1-8 keV energy band, in the first and second pointing respectively.

Source \#1 is by far the most intense and coincides with the source detected in the \textit{Swift}/XRT image (Landi et al. 2010), although the XRT uncertainty was not sufficient to pinpoint a unique counterpart.
A simple power law model with galactic absorption does not give an acceptable fit ($\chi^2/$d.o.f.=554/82), but the 
addition of an intrinsic absorber reduces the $\chi^2/$d.o.f. to $105/81$. This model 
gives an acceptable fit to the data, although a soft excess is also visible below 3 keV
in the residuals obtained with respect to the fitted model. Adding a thermal component (modeled in XSPEC with
BREMSS) resulted in a substantial fit improvement, reducing the $\chi^2/$d.o.f. from $105/81$ to $65/79$.
This corresponds to a low F-test chance probability of $6\times 10^{-9}$.
With this two-component model we obtain the following parameters:
a very high column density  $N_{\rm H}=(30_{-5}^{+3})\times10^{22}$cm$^{-2}$, a spectral index $\Gamma=1.2\pm0.4$,
a temperature of the thermal component of $0.21\pm0.07$ keV and unabsorbed fluxes of
$\sim1.2\times10^{-12}$ erg~cm$^{-2}$~s$^{-1}$  in 0.3-10 keV energy band.
The extrapolated high energy flux ($\sim5.7\times10^{-12}$ erg~cm$^{-2}$~s$^{-1}$
 in the 20-40 keV energy band) is  in full agreement with the IBIS detection ($\sim4.7\times10^{-12}$ erg~cm$^{-2}$~s$^{-1}$ in 20--40 keV energy range). The ACIS spectrum of this source is shown in figure 3.

Source \#2 has a  softer continuum ($\Gamma$$\sim$2.4) and a lower unabsorbed 0.3--10 keV
flux ($\sim0.4\times10^{-12}$ erg~cm$^{-2}$~s$^{-1}$) than Source \#1; its extrapolated high energy flux
is a factor of $\sim$50 lower than that reported by Bird et al. (2010) and so it cannot be considered a likely counterpart of the IBIS object. Similarly, all the other objects in Table 3 are unlikely associations.

Sources \#3 and \#4 are both much weaker (see Table 3) and so very unlikely to be the counterparts of the IBIS source.

This leads us to discard the association between the YSO and the IBIS object proposed
by Leyder, Walter \& Rauw (2008) and to conclude that Source \#1 is the only possible counterpart.  
We do not see any variability in the source spectral parameters between ACIS/\textit{Chandra} and XRT/\textit{Swift} observations, indicative of a persistent and possible low variability source.
Contrary to the XRT observation, the \textit{Chandra} restricted position allows us to pinpoint the infrared counterpart
of the source to the object 2MASSJ10445192-6025115.
This source has magnitudes in the J, H and K bands of 15.308, 14.967 and 13.977 respectively,
which combined with the lack of an optical detection in the USNO B1.0 catalogue implies a quite red object. 
The limit on R is estimated to be around 17 magnitude in the region surrounding source \#1, implying that R-K is $\ge$ 3.
The very high column density  ($N_{\rm H}=\sim2\times10^{23}$cm$^{-2}$) is compatible with the source reddening.
Only infrared follow-up observations with spectroscopic capability
will eventually provide information on the nature of this intriguing object.

\subsection{IGR J12288+0052}
This source is a new \textit{INTEGRAL} detection reported as a possible AGN in the 4$^{th}$ IBIS catalog (Bird et al. 2010),
with a weak 20-40 keV flux corresponding to $\sim4.5\times10^{-12}$ erg~cm$^{-2}$~s$^{-1}$. Indeed,
within the IBIS error box there are two AGN reported in the Catalogue of Quasars and Active Galactic Nuclei by Veron-Cetty and Veron (12th edition): the type 2 QSO SDSS 122845.74+005018.7 (RA=12 28 45.7, Dec=+00 50 18)  at z=0.57 and  the QSO 2QZJ122859+0054 (RA=12 28 59.1 Dec=+00 54 18) at z=1.2. This sky region is however full of other galaxies, some of which are also listed as quasar candidates selected from the photometric imaging data of the Sloan Digital Sky Survey (Richards et al 2009).
Within the IBIS error box at 99\% confidence level we detect various X-ray emitters and
five objects in the 2-8 keV energy range (see Figure 4); only one of these 5 is still visible above 4 keV and is located inside the 90$\%$  IBIS error box.\\
Source \#1 has a counterpart in the SDSS (SDSS J122833.46+005139.4), which is still unclassified;
Source \#2 is coincident with a QSO candidate in the SDSS (SDSS J122903.62+005359.5); 
Source \#3 is instead coincident with the type 2 QSO at z=0.57 and is the only source still visible at high energies.
Source \#4 is a galaxy listed in the SDSS as SDSS J122828.52+004827.8.
Finaly, Source \#5 has  no obvious counterparts in the various archives.

The spectrum of source \#3 (the brightest and hardest of all the objects detected),  although of poor quality,
is compatible with an AGN  canonical power law absorbed by a column density of $\sim$ 3 $\times10^{22}$cm$^{-2}$.
 The extrapolated flux in the 20-40 keV energy range is the highest among the 5 objects ($0.2\times10^{-12}$erg~cm$^{-2}$~s$^{-1}$)
although it is still much lower than that reported in the 4th IBIS catalogue ($\sim4\times10^{-12}$erg~cm$^{-2}$~s$^{-1}$):  
this suggests that either the source is variable or extremely absorbed and hence Compton thick. This source has been discussed by Vignali et al. (2010), who exclude on the basis of combinated optical, mid-IR and soft X-ray information that it could be a Compton thick AGN; this leaves the flux variability as a likely explanation for the ACIS and IBIS mismatch. Alternatively Source
\#3 could not be the real counterpart of the {\it{INTEGRAL}} detection, given the fact that the entire IBIS error box is not covered by {\it{CHANDRA}} observation.
Nevertheless the type II object SDSS 122845.74+005018.7 remains a potentially interesting counterpart candidate which deserves further studies.
It is reported as a radio source of 3.14 mJy flus dendity at 1.4 GHz, as an IRAS object with a $60\mu$m detetion of 120 mJy and has B and R magnitudes of 22.2 and 20.6 respectively. Only deeper and simultaneus broad band X-ray observation as possible with Suzaku could provide more detailed information
and confirmation of the source association to the {\it{INTEGRAL}} detection.

\begin{figure*}[h]
\includegraphics[angle=-90, width=0.8\linewidth]{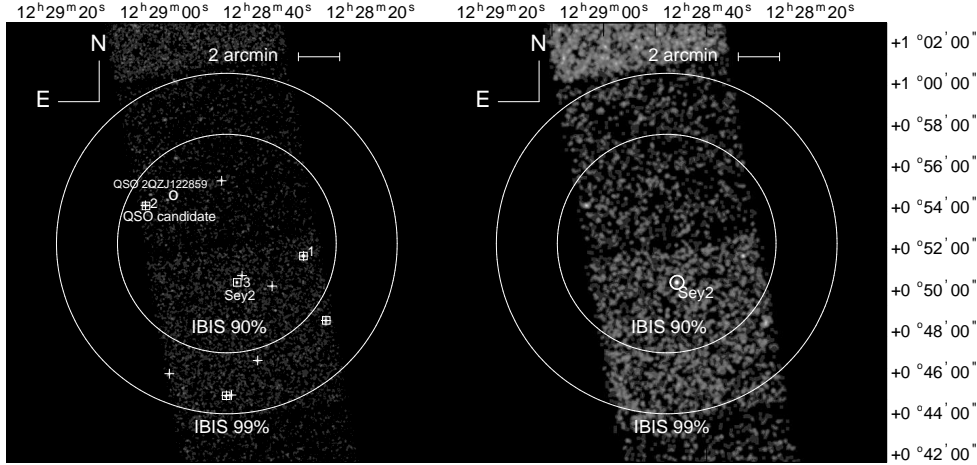}
\figcaption{
Left: ACIS 0.5-8 keV image of the region surrounding IGRJ12288+0052.  A Gaussian smoothing was applied to the count distribution with a width of 2 pixels.  The large circles represent the IBIS position and uncertainties expressed as 90\% and 99\%  confidence circles (as reported by Bird et al. 2010). Crosses indicate detections in the 0.5--2.0 keV band, squares detections in the 2.0-8.0 keV band and numbers indicate the position of the X-ray sources detected within the IBIS error box in the hard X-ray band (4--8 keV).  Right: ACIS 4--8 keV image of the region surrounding IGRJ12288+0052.  A Gaussian smoothing was applied to the counts distribution with a width of 4 pixels.  The large circles represent the IBIS position and uncertainties expressed as 90\% and 99\%  confidence circles (as reported by Bird et al. 2010). Circles represent sources detected in 4--8 keV energy band.}
\label{fig2b}
\end{figure*}

\begin{figure}[h]
\includegraphics[angle=-90, width=0.8\linewidth]{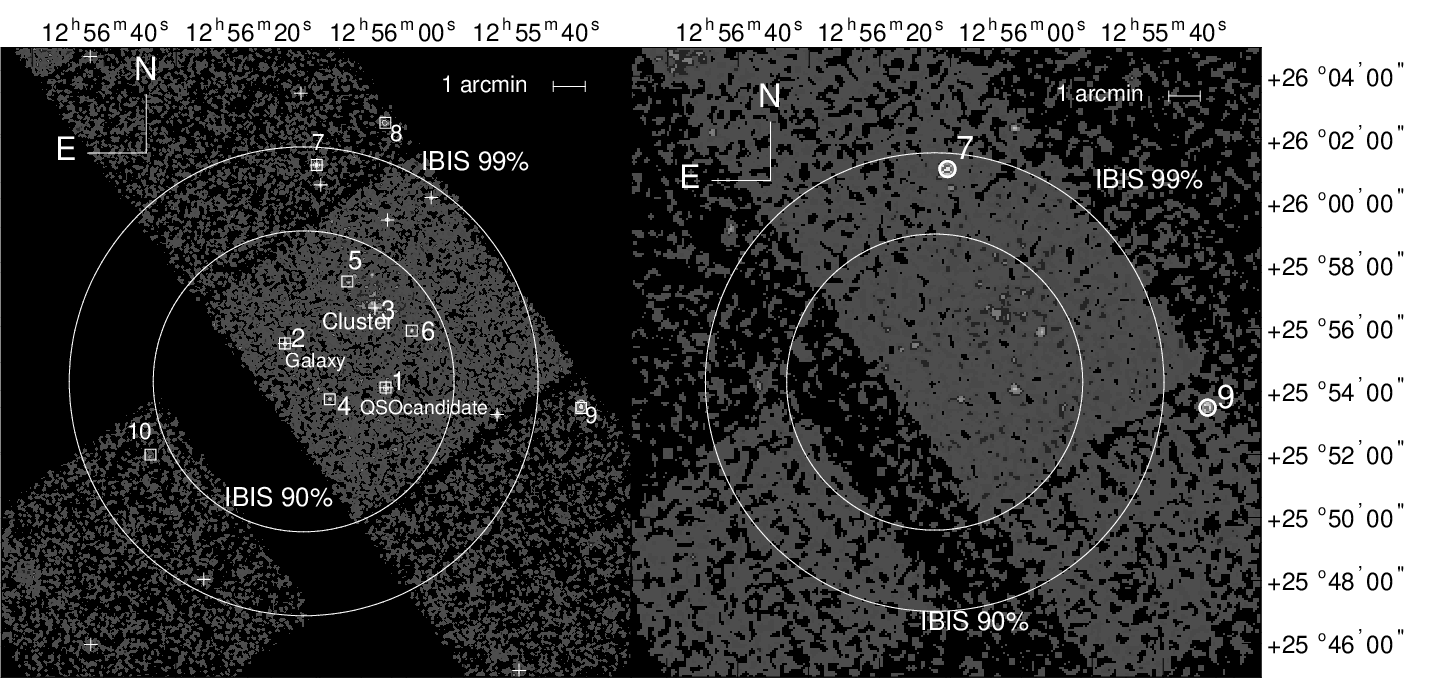}
\figcaption{
Left: ACIS 0.5-8 keV image of the region surrounding IGRJ12562+2554.  A Gaussian smoothing was applied to the counts distribution with a width of 2 pixels.  The large circles represent the IBIS position and uncertainties expressed as 90\% and 99\%  confidence circles (as reported by Bird et al. 2010). Crosses indicate detections in the 0.5-2.0 keV, squares detections in the 2.0--8.0 keV and numbers indicate the position of the X-ray sources detected within the IBIS error box in the hard X-ray band (4--8 keV).  Right: ACIS/CHANDRA and PN/XMM-Newton 4--8 keV image of the region surrounding IGRJ12562+2554.  A Gaussian smoothing was applied to the counts distribution with a width of 4 pixels.  The large circles represent the IBIS position and uncertainties expressed as 90\% and 99\%  confidence circles (as reported by Bird et al. 2010). Circles represent sources detected in 4--8 keV energy band.
Object \#1 is coincident with QSO candidate SDSS~J125600.78+255406.0,
object \#2 with the galaxy SDSS~J125614.76+255535.4 and object \#3 with the galaxy cluster WHL~J125602.3+255637.
}
\label{fig3a}
\end{figure}

\subsection{IGR J12562+2554}

This source is a new \textit{INTEGRAL} detection reported in the 4th IBIS catalog (Bird et al. 2010); it has
a 20-40 keV flux of 4.5$\times 10^{-12}$erg~cm$^{-2}$~s$^{-1}$  and is tentatively classified as a  cluster of galaxies.
Within the IBIS error box, an archival search
show that  the {\em {fossil}} group of galaxies RX J1256.0+2556 at z=0.232 is present.
Interactions  in a galaxy group cause large galaxies to spiral slowly towards the centre of the group, where they can merge to form a single giant central galaxy, which progressively swallows all its neighbours. 
If this process runs to completion, and no new galaxy falls into the group, then the result is an object dubbed a 'fossil group', in which an elliptical galaxy sits at the centre of a hot gas halo emitting X-rays.
Therefore the IBIS source could be associated with the cluster itself, the elliptical galaxy at its centre or any other member of the group.

Within the IBIS positional uncertainty, the \textit{Chandra} image in figure 5 shows the presence of a few soft sources and nine objects detected in the 2-8 keV band
(although two of these are located outside the 99$\%$ IBIS error box); only two objects are detected in the 4-8 keV energy band  and of these one is an unlike association as it falls outside the larger IBIS error circle.

The fossil group (source \#3), also serendipitously detected by XMM as 2XMM J125602.2+255636 (Watson et al. 2009) 
with a flux of $3\times 10^{-14}$erg~cm$^{-2}$~s$^{-1}$ (in the 0.3-10 keV energy band) is only seen below 2 keV. Recent analysis of the X-ray data from this system
gives a low gas temperature of $2.4-2.6$ keV using both XMM (Jeltema et al. 2006) and \textit{Chandra} (Jones et al. 2003)
data; we find a similar temperature indicative of a soft and weak X-ray source (see Table 3) which is  unlikely to be the counterpart of the IBIS  object.

Object \#1 is coincident with SDSS~J125600.78+255406.0, also serendipitously detected by XMM (2XMM J125600.8+255406,  Watson et al. 2009),
with a flux of $1\times 10^{-13}$erg~cm$^{-2}$~s$^{-1}$ (in the 0.3-10 keV energy band) and also listed as a \textit{Chandra} QSO candidate 
(CXOXA J125600.7+255406, Ptak \& Griffiths 2003).

Object \#2 is possibly the galaxy SDSS J125614.76+255535.4, also reported as 
2XMM J125614.8+255529 (Watson et al. 2009), with a
flux of $6\times 10^{-14}$erg~cm$^{-2}$~s$^{-1}$ (in the 0.3-10 keV energy band).

A number of sources seen in the 2-8 keV band (\#4,5,6,10) have no obvious identifications while a couple of objects (\#7,8) have a counterpart in the SDSS.

Object \#7 is the brighest and hardest source detected by {\it{CHANDRA}} within the IBIS positional uncertainty,
although it is located at the border of the 99$\%$ error circle and has a low X-ray flux of $2.5\times10^{-13}$erg~cm$^{-2}$~s$^{-1}$ in 0.3-10 keV band. It is listed as a quasar candidate at z=1.2
in the  SDSS NBCKDE Catalog of Photometrically Selected Quasar Candidates (Richards et al. 2009).
The source spectrum is a power law with $\Gamma\sim$1.7 and galactic $N_H$;
the {\it{CHANDRA}} spectrum extrapolated to the 20-100 keV band gives $\sim 0.5\times 10^{-12}$erg~cm$^{-2}$~s$^{-1}$, a factor of 9 lower than the IBIS detection.

Since the IBIS error circle is not fully covered by \textit{Chandra} we
analysed the  \textit{XMM}/EPIC image (observation ID 0012850201), available in the XMM archive\\
(http://xmm.esa.es/xsa/index.shtml) in the 4--8 keV energy band. This study shows that no other X-ray
sources are detected within the IBIS error circle, leaving the association with object \#7 the only possibility at this stage.
To reconcile the source spectrum with the IBIS detection, we need to assume that either object \#7
is variable or again heavly absorbed  than observed.
In this last case, {\it{INTEGRAL}} may have detected a new type 2 quasar at high redshift.
The fact that {\it{CHANDRA}} spectrum measure a column density in excess of the galactic value indicates flux variability as the origin
of the mismatch. The source is reported in the USNO B1.0  catalogue as having B and R in
the range magnitudes 19.9-20.8 and 19.2-19.8, respectively, which supports a variable source.
In this case too, a dedicated broad band X-ray observation as possible with Suzaku could provide the necessary
insight into the nature of this source and of its association with the {\it{INTEGRAL}} object.

\subsection{IGR J14193-6049}
This source is a very weak new \textit{INTEGRAL} detection  reported in the 4th IBIS catalog (Bird et al. 2010)
with flux of 3.8$\times 10^{-12}$erg~cm$^{-2}$~s$^{-1}$ in 20-40 keV energy band and located 
at RA(J2000)=214.821 and DEC(J2000)=-60.801 with a positional uncertainty 
of 4.3$^{\prime}$ (90\% confidence level).
Detailed spectral analysis of this sky region was performed by Ng et al. (2005).
These authors listed a number of X-ray sources, some of which are related to the so called Kookabarra complex region, which has two Supernovae/pulsar wind nebulae emitting nearby
and also detected at TeV energies (Aharonian et al. 2006).
Within the IBIS error box at 99\% confidence level, Ng et al. (2005) reported
four sources. Two are bright and soft X-ray objects (Star1: RA=14 19 11.75, DEC=-60 49 33.8; Star2: RA=14 19 31.65, DEC=-60 46 21.3), clearly identified with bright field stars. The
faintness in X-rays and soft spectra (few counts above 2 keV) of both objects
suggest that their association with IGR J14193-6049 is very unlikely.
The third source is called Sr3 (RA=14 18 37.83, DEC=-60 45 01.1): Ng et al. 2005 suggested that this object is an active galactic nuclei 
also detected in the radio band as a flat spectrum, variable point source with a radio flux of a few mJy.
With an extrapolated 20-40 keV flux  of $\sim0.3 \times 10^{-12}$ erg~cm$^{-2}$~s$^{-1}$ (using the spectral parameters reported by Ng et al. (2005)),
this source is however too weak to be the  counterpart of the IBIS source. However if we enlarge the error box to the 99\% confidence level, we find a new detection in PSR J1420-6048, a pulsar which also possesses a sorrounding nebula (Roberts et al. 2001):
this source is the brightest and hardest of all the sources detected by {\it{CHANDRA}} in the IBIS error box.
Figure 6 shows the ACIS image together with the H.E.S.S. (Aharonian et al . 2006) and IBIS error circles, which clearly suggests that both are detecting the same object.
Using the total flux of PSR J1420-6048 (PSR plus PWN) and a spectral index of $\sim$ 1.6 as reported by Possenti et al. (2002) we obtained an extrapolated flux
of $\sim4.3 \times 10^{-12}$erg~cm$^{-2}$~s$^{-1}$
in the 20-40 keV energy band, which is in very good agreement with the IBIS detection reported by Bird et al. (2010). We therefore conclude that  PSR J1420-6048 (pulsar and nebula)  is the counterpart of IBIS source.
Pulsars and their  nebulae are strong emitters in the keV and TeV band: indeed 13 such systems  
have now been detected  by {\it{INTEGRAL}} in the 20-100 keV energy band and of these 9 have a counterpart in H.E.S.S., strengthening the association of PSR J1420-6048 with IGR J14193-6049.

\begin{figure}[h]
\includegraphics[angle=-90, width=0.8\linewidth]{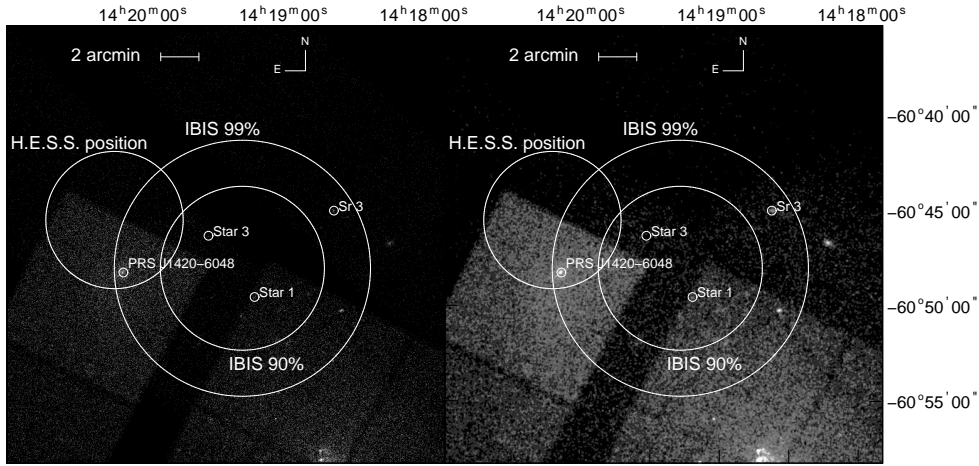}
\figcaption{
Left: ACIS 0.5-8 keV image of the region surrounding IGRJ14193-6048.   The large circles represent the IBIS position and uncertainties expressed as 90\% and 99\% confidence circles (as reported by Bird et al. 2010). The X-ray sources are labelled as reported in Ng et al (2005). Small circles indicate the
H.E.S.S. position of the pulsar with nebula PSR J1420-6048 (Aharonian et al. 2006).   Right: ACIS/CHANDRA and MOS/XMM-Newton 4--8 keV images of the region surrounding IGRJ14193-6048.  A Gaussian smoothing was applied to the counts distribution with a width of 2 pixels.  The large circles represent the IBIS position and uncertainties expressed as 90\% and 99\%  confidence circles (as reported by Bird et al. 2010).
The X-ray sources are labelled as reported in Ng et al (2005). Small circles indicate the
H.E.S.S. position of the pulsar with nebula PSR J1420-6048 (Aharonian et al. 2006).
}
\label{fig14}
\end{figure}

\begin{figure}[h]
\includegraphics[angle=-90, width=0.8\linewidth]{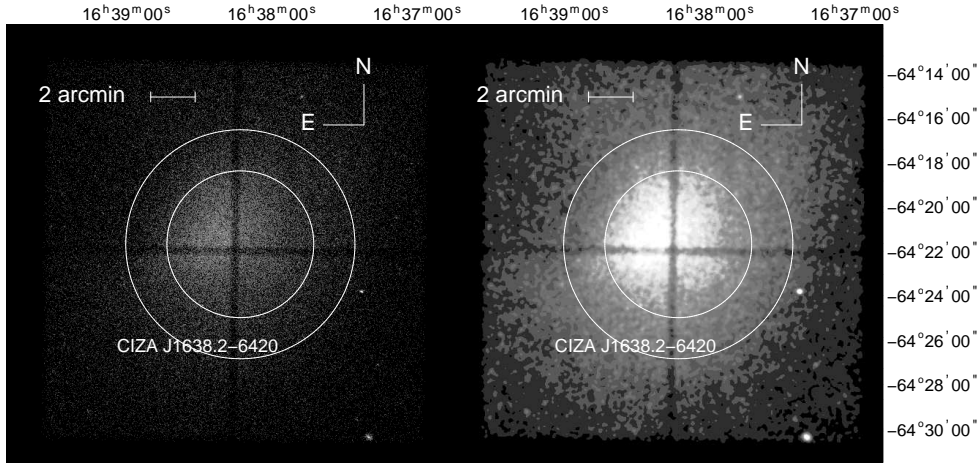}
\figcaption{
Left: ACIS 0.5-8 keV image of the region surrounding  IGRJ16377-6423.  The large circles represent the IBIS position and uncertainties expressed as 90\% and 99\%  confidence circles (as reported by Bird et al. 2010).  Right: ACIS 4--8 keV image of the region surrounding IGRJ16377-6423.  A Gaussian smoothing was applied to the counts distribution with a width of 4 pixels.  The large circles represent the IBIS position and uncertainties expressed as 90\% and 99\%  confidence circles (as reported by Bird et al. 2010). }
\label{fig3b}
\end{figure}

\subsection{IGR J16377-6423}
IGR J16377-6423 is a source detected by \textit{INTEGRAL} in the 17-30keV energy band and tentatively associated by Bird et al. (2010)
to a cluster of galaxies. This was mainly due to the fact that ROSAT observations clearly indicate the presence of a bright galaxy cluster CIZA J1638.2-6420 (Stephen et al. 2006).
It was not clear if {\it{INTEGRAL}}
had detected the cluster itself or one of its members. Within the IBIS error box we find an extended and bright X-ray source clearly identified with the galaxy cluster even at high energies (see Figure 7).
The analysis of the spectrum was performed by Snowden et al. (2008), 
showing that this cluster has a thermal spectrum with a very high plasma temperature in the range $8-13$ keV  and a flux of $0.1-3.3\times10^{-12}$erg~cm$^{-2}$s$^{-1}$ in the 0.3-10 keV energy band.
Such high values of gas temperature could provide detectable emission in the IBIS waveband, making the association with the \textit{INTEGRAL} source plausible.
In fact using the fit parameters obtained  by Snowden et al. (2008), we derive an extrapolated 20-40 keV flux of  $\sim3\times10^{-12}$erg~cm$^{-2}$s$^{-1}$,
comparable with the one reported by Bird et al. (2010) ($\sim7\times10^{-12}$erg~cm$^{-2}$s$^{-1}$).
The 4{th} \textit{INTEGRAL} survey  (Bird et al. 2010) reports only four cluster of galaxies  out of 723 sources detected;
all four of them are bright and with a high temperature.  This suggests that the high energy emission is directly linked to 
this cluster property.\\

\section[]{Conclusions}
One of the more difficult task associated with the observation of the high energy sky, though with non focussing optics, is associated to the identification of a substantial fraction of the newly discovered sources. In fact, in spite of the forward step achieved in the imaging capability of the  new generation high energy instruments (e.g. SWIFT and {\it{INTEGRAL}} in the hard X-ray range, FERMI and AGILE in the gamma-ray domain and Cherenkov ground telescope at TeV energies) optical identification remain a very difficult task. This task becomes almost impossible whenever timing informations or evident source variability are not detected at high statistical significance at different wavelenghts (e.g. GRBs, Pulsar, AGNs etc). In particular, for weak sources the optical and/or IR identification remain a challenge if arcsec X-Ray counterpart position are not available.

To overcome this problem a robust programme has started in parallel with the different {\it{INTEGRAL}}/IBIS survey production to obtain, whenever possible, X-Ray fields containing counterpart for the high energy unknown detected objects. This aproach has been successful using archival data from ROSAT, and TOO dedicated observation and archival data from XRT, XMM-Newton and {\it{CHANDRA}}. This data were essential to trigger a robust programme of ground based optical observations and identification that has been  very succesfull so far (Masetti et al. 2009 and references therein ).

In this paper we have used a cross correlation between the more recent Integral IBIS survey results and the {\it{CHANDRA}} archival data to disentangle the nature of still unclassified 5 objects. So far, we have identified candidate soft X-ray counterparts in four cases and extracted their soft X-ray spectra. Moreover, the present data have provided further basic information into the nature of these sources and better than 1 arcsec refined positions essential for the planned follow-up observations.

\section*{Acknowledgments}

The authors acknowledge the ASI financial support
via ASI-INAF contract I/008/07/0.
We thank the reviewer for his/her thorough review and highly appreciate the comments and
suggestions, which significantly contributed to improving the quality of the publication.
This research has made use of data obtained from the SIMBAD
database operated at CDS, Strasbourg, France; the
High Energy Astrophysics Science Archive Research Center
(HEASARC), provided by NASAs Goddard Space Flight
Center NASA/IPAC Extragalactic Database (NED). We
also acknowledge the use of public data from the \textit{Chandra} data
archive.

\scriptsize

\begin{deluxetable}{ccccccccl}
\rotate
\tabletypesize{\scriptsize} 
\tablecaption{INTEGRAL/IBIS position of the 5 selected sources (from Bird et al. 2010) and locations of the sources detected by {\it{CHANDRA}}/ACIS
with their respective counterparts.  The positional uncertainties are 1-$\sigma$ statistical errors as computed by the  {\ttfamily wavdetect} software.
Sources with asterisks are detected in the hard band (4-8 keV) but they are outside of the IBIS error box at 99\% confidence level. }
\tablewidth{0pt}
\tablehead{ 
\colhead{\textit{\#}} &
\colhead{RA J2000} & 
\colhead{Dec J2000} & 
\colhead{RA Error} &
\colhead{DEC Errr } &
\colhead{Count Rate  } &
\colhead{Count Rate} &
\colhead{Count Rate} &
\colhead{Counterparts}\\
\colhead{} &
\colhead{} & 
\colhead{} & 
\colhead{($\prime\prime$)} &
\colhead{ ($\prime\prime$)} &
\colhead{($4-8 keV$) } &
\colhead{ ($2-8 keV$)} &
\colhead{ ($0.5-2 keV$)} &
\colhead{}
}

\startdata 
\hline
\\
\multicolumn{9}{c}{IGR J10447-6027, (R.A.(J2000)=161.155, Dec(J2000)=-60.423, error radius=4.1$^{\prime}$)
}\\
\\
{\bf{1}}&{\bf{161.2162}}   & {\bf{  -60.4200}}  &    0.1  & 0.1 &  486  $ \pm $  22 &589  $ \pm $  25&...&2MASS10445192-6025115\tablenotemark{a}\\  
2&161.1060   &   -60.3850  &    0.2 &  0.1 &    66$\pm$8   & 140 $\pm$12  &17$\pm$4&\\
3&161.2836   &   -60.4545  &   1.0  &  0.8 &  ...     &  20$\pm$5      & ...&...\\
4&161.0756  &    -60.3692  &    1.9 &  0.6 &  ... & 33$\pm$6&...&...  \\
5$^*$&160.8617  &    -60.4396  &    1.2 &  0.6 &  48 $ \pm $7 & 64$\pm$9&...&  ...\\
6&161.2906&-60.4615&0.8&0.5&...&64$\pm$9&...&...\\
7&161.0879&-60.5007&0.6&0.5&...&16$\pm$4&...&...\\
8&161.1899&-60.3955&0.8&0.4&...&19$\pm$4&...&...\\
9&161.2022&-60.3953&0.7&0.5&...&18$\pm$4&...&...\\
10&161.0314&-60.3422&1.4&0.7&...&21$\pm$5&...&...\\
11&161.2373&-60.5046&1.0&1.2&...&22$\pm$5&...&...\\
12&161.0502    &    -60.3420   &     2.2    &   0.5  &...&21$\pm$4&...                   &...\\
13&161.2696    &    -60.4523  &      0.71    &0.4    &...&...& 16  $\pm$    4          &...             \\
14&161.2117    &    -60.3422  &      0.83    &0.7    &...&...& 20  $\pm$    5          &...          \\
15&161.2138    &    -60.4354  &      0.91    &0.2    &...&...& 16  $\pm$    4          &...           \\
16&161.0611    &    -60.3923  &      0.82    &0.5    &...&...& 17$\pm$      4          &...             \\
17&161.0723    &    -60.4975  &      1.1     &0.5    & ...&...&28$\pm$      6          & ...       \\
18&161.0975    &    -60.4759  &      0.53    &0.5    &...&...& 29  $\pm$    6          &...          \\
19&161.1461    &    -60.3321  &      0.75    &0.3    &...&...& 36  $\pm$    6          &...       \\
20&161.336     &    -60.437  &        0.48  &  0.2  &  ...&...& 44$\pm$      7        &  ...      \\
21&161.2073    &    -60.3564  &      0.48    &0.3    &...&...& 44  $\pm$    7          &...          \\
\hline
\\
\multicolumn{9}{c}{IGR J12288+0052, (R.A.(J2000)=187.199, Dec(J2000)=0.870, error radius=5.3$^{\prime}$)
}\\
\\
1& 187.1370 & 0.8599&  0.1& 0.2&...&38$\pm$4&90$\pm$10&SDSS J122833.46+005139.4 \tablenotemark{h}\\
2& 187.2647&  0.9005&  1.0& 0.6&...& 21$\pm$5&20$\pm$5&SDSSJ122903.62+0053 \tablenotemark{d}\\
&&&&&&&&QSO Candidate\\
{\bf{3}}& {\bf{187.1906}}& {\bf{0.8386}}&  0.2& 0.2&16$\pm$4& 45$\pm$7&...&SDSS J122845.74+005018.7   \tablenotemark{e}\\
&&&&&&&&Seyfert Type 2\tablenotemark{f}\\ 
4&187.1188  &0.8078    &0.5   &1.1   &...& 13$\pm$3 & 27$\pm$5&  SDSS J122828.52+004827.8\\
5&187.1993  &0.7472    &0.4   &0.4  &...& 20$\pm$5&22$\pm$5&...\\
6&187.2032    &      0.9208   &       0.5   &  0.3   &... &...  & 18$\pm$4&...\\
7&187.1741 &       0.7755  &        0.6  &   0.4  &... & ... &  23$\pm$5&...\\
8&187.1953 &       0.7476 &        0.3   &  0.4   &... & ... & 24$\pm$5 &...\\
\hline
\\
\multicolumn{9}{c}{IGR J12562+2554, (R.A.(J2000)=194.051, Dec(J2000)=25.905, error radius=4.7$^{\prime}$)
}\\
\\
1& 194.0032& 25.9019& 0.9& 0.9& ... & 39 $\pm$7 & 96 $\pm$10 &CXOXA J125600.7+255406\tablenotemark{g}\\
&&&&&&&&SDSSJ125600.78+255405.9\tablenotemark{d}\\
&&&&&&&&QSO Candidate\\
2& 194.0619& 25.9247& 0.7& 0.4&...& 16 $\pm$4&48 $\pm$7&SDSSJ125614.76+255535.4\tablenotemark{h}\\
&&&&&&&&Galaxy\\
3& 194.0103& 25.9445& 0.4& 0.4&...&...& 130 $\pm$25 &WHL~J125602.3+255637\tablenotemark{i}\\
&&&&&&&&Cluster\\
 4  &  194.0359  &       25.8956    &      0.7 &     0.7   &   ...&   29 $\pm$      6          & ...  &      ....\\
 5  &  194.0258  &       25.9570    &      1   &      0.5  &...& 19$\pm$ 5     &     ...  &    ...      \\
 6  &  193.9885  &       25.9314    &      0.4 &     0.4   &... &20 $\pm$      5   &     ...    &   ...        \\
{\bf{7}}&  {\bf{194.0435}}  & {\bf{ 26.0177}}    &      0.1 &     0.1   & 55$\pm$8 & 170$\pm$13   &     393$\pm$20       &    SDSS J125610.42+260103.5 \tablenotemark{h}\\
 8  &  194.0037  &       26.0396    &      0.4 &     0.4 &... &          35 $\pm$      6&     &     SDSS J125600.92+260222.5 \tablenotemark{h}\\
 9$^*$  &  193.8904  &       25.8911    &      0.3 &     0.3 &     18$\pm$ 4  &  76  $\pm$     9&     &    V* IN Com              \\
10&  194.1397  &      25.8668     &    0.9  &   1  &... & 16  $\pm$          4 &    ...       &...                 \\
\hline
\\
\multicolumn{9}{c}{IGR J16377-6423, (R.A.(J2000)=249.553, Dec(J2000)=-64.362, error radius=3.3$^{\prime}$)
}\\
\\
 {\bf{1}}&  {\bf{249.5711}}&  {\bf{-64.3570}}& 0.9& 0.4& 6682 $\pm$268&2222 $\pm$158& 4383 $\pm$219&CIZA J1638.2-6420\tablenotemark{j}\\
\hline
\hline
\enddata

\tablenotetext{a}{Two Micron All Sky Survey, Skrutskie et al. 2006. } 
\tablenotetext{b}{The Guide Star Catalogue, Version 2.3.2,  Lasker B., Lattanzi M.G., McLean B.J., et al. 2008.} 
\tablenotetext{c}{First DENIS I-band Extragalactic Catalog, Vauglin I. et al. 1999.} 
\tablenotetext{d}{Sloan Digital Sky Survey NBC Quasar Candidate Catalog,  Richards G.T. et al. 2004.} 
\tablenotetext{e}{The Sloan Digital Sky Survey quasar catalog, Schneider D.P.,  et al. 2007.} 
\tablenotetext{f}{Veron Catalog of Quasars \& AGN, 12th Edition Veron et al. 2006.}
\tablenotetext{g}{Chandra XAssist Source List,   Ptak A. \& Griffiths R., 2003}
\tablenotetext{h}{The SDSS Photometric Catalog, Release 7, Adelman-McCarthy et al.  2009.}
\tablenotetext{i}{Sloan Digital Sky Survey DR6 Galaxy Clusters Catalog, Wen Z.L., Han J.L., Liu F.S. 2009.}
\tablenotetext{j}{CIZA catalog, Ebeling et al. 2002 .}
\end{deluxetable}

\begin{center}
\begin{table*}[t]
\scriptsize
\caption{{\em {\it{CHANDRA}}} ACIS spectral analysis results. A $N_H$ fixed to the Galactic column density was included in the fit. Error are given at $90\%$ confidence level for one parameter of interest ($\Delta\chi^2=2.71$). Unabsorbed (0.3--10) keV and (20-40) keV fluxes are reported in units of $10^{-12}$ erg~cm$^{-2}$~s$^{-1}$. }
\label{tab:spectra}
\scriptsize
\begin{tabular}{cccccccc} \hline \hline
Sources &off-axis &model & $N_{\rm H}$  & $\Gamma$ & X-ray Flux & Extrapolated Flux &$\chi^2/d.o.f.$\\
 & & &  ($\times 10^{22}$ cm$^{-2}$) & & $0.3-10 keV$&$20-40 keV$&\\
\hline \hline

\multicolumn{8}{c}{IGR J10447-6027 (unidentified, $N_{\rm Hgal}=1.3\times10^{22}$cm$^{-2}$)
}\\
\\
1&0.9$^\prime$&power law &  $21\pm3$  &$1.0^{+0.3}_{-0.6}$    & $1.7\pm0.3$   &5.7 &105/81  \\
2&4.1$^\prime$&power law & $4^{+3}_{-2}$    & $2.2^{+1.3}_{-0.7}$   &  $0.4^{+2.0}_{-0.3}$  &$<$1.1& 19/14   \\
\\
\hline
\\
\multicolumn{8}{c}{IGR J12288+0052, (AGN?, $N_{\rm Hgal}=1.9\times10^{20}$cm$^{-2}$)
}\\
\\
3&0.4$^\prime$& power law&$2.5^{+6.5}_{-3.1}$    &$1.6\pm0.8$    &  $0.3^{+3.0}_{-0.2}$    &0.2&  6/7     \\
\\
\hline
\\
\multicolumn{8}{c}{IGR J12562+2554, (Cluster?, $N_{\rm Hgal}=7.6\times10^{19}$cm$^{-2}$)
}\\
\\
7&4.5$^\prime$&power law&   ...  &$1.7\pm0.1$    & $0.25\pm0.04$  \epsscale{0.85}&0.3& 32/30  \\

\hline\hline
\end{tabular}
\end{table*}
\end{center}

%
%

%

\end{document}